\documentclass[useAMS,usenatbib]{mn2e}

\usepackage{enumerate}
\usepackage{multirow}
\usepackage{graphicx}
\usepackage{natbib}
\usepackage{longtable}
\usepackage{rotating}
\usepackage[usenames,dvipsnames]{color}

\title[Molecular hydrogen formation]{Effect of surface H$_2$ on molecular hydrogen formation on interstellar grains}
\author[Zhao et al.]{Gang Zhao $^{1}$, Qiang Chang$^{1}$\thanks{email:{changqiang@sdut.edu.cn}}, 
                     Xia Zhang$^{2}$, Donghui Quan$^{3}$,
		     Yong Zhang $^{4}$, Xiao-Hu Li$^{2}$ \\
$^{1}$School of Physics and Optoeletronic Engineering, Shandong University of Technology, 
     Zibo, 255000, China\\
$^{2}$Xinjiang Astronomical Observatory, Chinese Academy of Sciences, 150 Science 1-Street, Urumqi 830011, China\\
$^{3}$Department of Chemistry, Eastern Kentucky University, Richmond, KY, USA\\
$^{4}$School of Physics \& Astronomy, Sun Yat-Sen University, Zhuhai 519082, China
}

\begin{document}

\maketitle

\label{firstpage}

\begin{abstract}
We investigate how the existence of hydrogen molecules on grain surfaces may affect H$_2$ formation 
efficiency in diffuse and translucent clouds.
Hydrogen molecules are able to reduce the desorption energy of H atoms on grain surfaces in models. 
The detailed microscopic Monte Carlo method is used to perform model simulations. 
We found that the impact of the existence of H$_2$ on H$_2$ formation efficiency strongly 
depends on the diffusion barriers of H$_2$ on grain surfaces. 
Diffuse cloud models that do not consider surface H$_2$ predict that 
H atom recombination efficiency is above 0.5 over a grain temperature (T) range 10 K and 14 K.
The adopted H$_2$ diffusion barriers in diffuse cloud models that consider surface H$_2$ 
are 80$\%$ H$_2$ desorption energies so that H$_2$ can be trapped in stronger binding sites.
Depending on model parameters, these diffuse cloud models 
predict that the recombination efficiency is between nearly 0 and 0.5 
at 10 K $\leq$ T $\leq$ 14 K. 
Translucent cloud model results show that H$_2$ formation efficiency is not affected 
by the existence of surface H$_2$ if the adopted average H$_2$ diffusion barrier on grain surfaces is low (194 K)
so that H$_2$ can diffuse rapidly on grain surfaces. 
However, the recombination efficiency can drop to below 0.002 at T $\geq$ 10 K 
if higher average H$_2$ diffusion barrier is used (255 K) in translucent cloud models.

\end{abstract}



\section{Introduction}\label{sec:intro}
Molecular hydrogen is the most abundant molecules in astronomical sources. Moreover, gas-phase H$_2$ 
is crucial for the subsequent formation of 
the molecular ion H$^{+}_3$, which triggers the rich gas-phase interstellar chemistry \citep{Williams1998,Herbst2001}. 
Therefore, the H$_2$ formation mechanism has been an interesting and important research topic since 
its discovery. Although almost all interstellar molecules were believed to be formed in the gas phase in
astrochemical models, H$_2$ formation in the gas phase has been found to be inefficient while 
interstellar dust grains can play an important role in increasing the efficiency with which H atoms convert
to H$_2$ molecules~\citep{Gould1963}.

However, it is not straightforward to explain H$_2$ formation in astronomical sources even 
when the catalytic roles of dust grains are introduced into models. 
Interstellar species are believed to be formed on cold grain surfaces via 
the so called Langmuir-Hinshelwood 
mechanism~\citep{Watson1972,Pickles1977,Hasegawa1992}. To form H$_2$, 
H atoms accrete on dust grains and then bind weakly with surfaces, which is known as 
physisorption. They can overcome the diffusion barrier and move on the grain 
surfaces via quantum tunneling or thermal hopping.
However, laboratory and theorectical studies showed that quantum tunneling does not contribute 
much to the mobility of H atoms on silicate, carbonaceous or solid amorphous water (ASW) 
surfaces~\citep{Pirronello1997,Pirronello1999,Katz1999,Nyman2021}.
If H atoms encounter other
H atoms, then H$_2$ molecules are formed. But
H atoms can also desorb and leave grain surfaces.
A hydrogen atom must reside on a grain
long enough to find a partner H atom to form H$_2$. 
As the dust temperature increases, the H atom desorption and diffusion rates also increase.
So the temperature of grain surfaces must be sufficiently low so that a H atom can encounter another one before it desorbs. 
On the other hand, the temperature of grain surfaces must be high enough so that H atoms can diffuse 
on the grain surface. 
The parameter that measures how strongly species are to bound to grain surfaces is called desorption
energy. 
It was found that if we assume a single H desorption energy in models, 
the dust temperature range over which efficient H$_2$ formation occurs 
is narrow (6-10 K for olivine grains)~\citep{Katz1999}.
Moreover, it was found that the highest dust temperature at which H$_2$ can be formed efficiently is less than 17 K~\citep{Katz1999}.
However, the grain surface temperature in the unshielded diffuse clouds, where hydrogen molecules are believed to be efficiently formed, 
is around 20 K~\citep{Li2001}. 

In order to make the dust temperature range over which efficient H$_2$ formation occurs wider,
H$_2$ formation models adopted at least two types of sites, the stronger and weaker binding sites, 
on grain surfaces~\citep{Hollenbach1971b,Chang2005,Cuppen2005,Iqbal2012}.
In these models, the stronger binding sites are able to hold H atoms on grain surfaces long enough 
so that other H atoms can encounter them to form H$_2$ before they desorb even when 
the dust temperatures are high. On the other hand, H atoms in the weaker binding sites 
can diffuse to encounter other H atoms even when the dust temperatures are low.
These models are able to significantly enlarge the dust temperature range over 
which efficient H$_2$ formation occurs.
The upper limit of dust temperatures at which the efficient conversion from H to H$_2$ occurs
depend on the strongest binding sites on grain surfaces. As the desorption energy 
of the strongest binding sites increases, this upper limit also increases.
The recent review by \citet{Wakelam2017a} further summarizes H$_2$ formation models, experiments and observations.

Because H$_2$ molecules cannot react with H atoms on grain surfaces, 
the impact of the existence of surface H$_2$ was not considered in earlier H$_2$ 
formation models~\citep{Hollenbach1971b,Biham2001,Chang2005,Cuppen2005,Iqbal2012}. 
However, studies using astrochemical models for the cold ($\sim10$ K) and high density
(H nucleus densities $>10^{12}$ cm$^{-3}$) sources showed that 
in order for the modeling results to be physical,
the H$_2$ desorption energies in binding sites occupied by H$_2$ must 
be much lower than that on water ice~\citep{Garrod2011,Hincelin2015}.
If a chemical model adopts a single H$_2$ desorption energy 
and the H$_2$ desorption energy on water ice is used in the model,
then gas-phase hydrogen molecules would be heavily depleted onto dust grains in the chemical model~\citep{Hincelin2015},
which has not been observed.  
To solve this problem, \citet{Hincelin2015} suggested the 
encounter desorption (ED) mechanism in their models. The ED mechanism assumes that
when a hydrogen molecule hops into a site occupied by another H$_2$ molecule, 
the desorption energy for the hopping H$_2$ molecule 
becomes the desorption energy of H$_2$ on H$_2$ substrate, which is much lower than that on water ice surfaces. 
So, surface H$_2$ can desorb much more efficiently even around 10 K if the ED mechanism 
is included in models.
As a result, models predict that 
the abundances of surface H$_2$ in cold and high density sources  
are only no more than a few monolayers~\citep{Hincelin2015}.
In a recent study, \citet{Chang2021} extended the H$_2$ ED to the H and H$_2$ ED,
which assumes that when H atoms encounter H$_2$ on grain surface,
their desorption energy becomes the H atom desorption energy on H$_2$ substrate, which is 
much lower than that on water ice surfaces. So, the H and H$_2$ ED mechanism enhances 
the desorption rate of surface H, thus,
it becomes more difficult to hydrogenate CO on grain surfaces.
So, the production of a few surface species such as methanol 
was strongly reduced if the model adopted the H and H$_2$ ED mechanism.

The desorption rate of H atoms should also increase if we consider H and H$_2$ ED mechanism in
H$_2$ formation models because the desorption energy of H on H$_2$ substrate is much lower than that 
on bare dust grain surfaces. On the other hand, the existence of H$_2$ on grain surfaces may 
affect H$_2$ formation in another way. The efficient H$_2$ formation at higher temperatures
relies on the existence of stronger binding sites on grain surfaces. If we consider physisorption only,
the stronger binding sites are those where surface species can see more horizontal 
neighbors, which are the upward step edges on rough surfaces~\citep{Cuppen2005}.  
Therefore, a stronger H binding site must also be a stronger H$_2$ binding site if we consider
physisorption only. After a hydrogen molecule occupies a stronger binding site, this site becomes
a weaker binding site for two reasons.  First, for the H atom in this site, 
the vertical neighbor below it becomes H$_2$. Secondly, the H atom may not see horizontal neighbors,
thus loses lateral bonds with it neighbors because a H$_2$ molecule fills the vacancy. We refer to \citet{Cuppen2005} 
for details of how horizontal and vertical neighbors affect the desorption energies of surface species.

The impact of the existence of surface deuterium molecules on D$_2$ formation 
has been studied in laboratory~\citep{Gavilan2012}. It was found that D$_2$ molecules formed 
more efficiently on silicate surfaces covered by D$_2$ than on bare silicate surfaces. 
\citet{Gavilan2012} argued that the reason was that D atoms diffuse more quickly 
on surfaces covered by D$_2$. However, the flux of D atoms in their experiments was more 
than seven orders of magnitude larger than
the H atom flux in diffuse clouds~\citep{Biham2001}.  
As the H atom accretion flux significantly decreases, the H atom population on grain surfaces 
also drops a lot. Because H$_2$ formation requires at least two H atoms on grain surfaces,
it is not clear if the existence of surface H$_2$ can still help to increase H$_2$ formation efficiency on grain surfaces.

In this paper, for the first time, we investigate H$_2$ formation by the recombination of 
physisorbed H atoms on interstellar dust 
grains using models that consider the existence of H$_2$ on grain surfaces.   
Although our major purpose is to study H$_2$ formation in diffuse clouds, we also investigate 
H$_2$ formation in translucent clouds, where
typically $25\%$ H nucleus are in the form of H atoms~\citep{Zuo2018}. These H atoms
will also be converted to H$_2$ as translucent clouds gradually become denser and form dense dark clouds~\citep{Zuo2018}.  

The organization of the paper are the follows.
Our models are explained in Section \ref{sec:models}
while the numerical method is introduced in Section \ref{sec:method}.
Our results are shown in Section \ref{sec:res}.
Finally, Section \ref{sec:sum} is the section 
for the discussions and conclusions.

\section{Models}\label{sec:models}
Based on the work by \citet{Biham2001}, we assume in our models that 
the H nucleus density of diffuse clouds is 10 cm$^{-3}$ 
while the gas-phase temperature is 100 K.
Although both olivine and carbon grains exist in diffuse clouds, we only study 
the H$_2$ formation efficiency on silicate-composed olivine grains in diffuse clouds.
The effect of H$_2$ existence on molecular hydrogen formation on carbon grains will be briefly 
discussed in Section \ref{sec:sum}.
Assuming the gas-phase species is composed of H atoms only, the H atom accretion flux 
is $f1 = 1.8\times 10^{-9}$ ML s$^{-1}$ in diffuse cloud models \citep{Biham2001}. 
Diffuse clouds are those in which more than $10 \%$ H nuclei have been converted to H$_2$~\citep{Wakelam2017a}, 
so, we assume $75 \%$ or  $50 \%$
H nuclei exist in the form of H atoms in diffuse cloud models. 
The purpose of this work is to study how the existence of surface H$_2$ may affect only the
physisorbed H atom recombination efficiency.
Therefore, H$_2$ formation scenarios involving chemisorption are ignored in models.

In translucent clouds, the gas temperature is 15-40 K, the H nucleus density is 
between 500 and 5000 cm$^{-3}$ while dust grains are believed to be covered
by water ice~\citep{Wakelam2017a}.
In translucent cloud models, we therefore assume that the H nucleus density is 
1000 cm$^{-3}$ and the gas temperature is 30 K. 
In a typical translucent cloud,  $75 \%$ of the H nuclei have been 
converted to H$_2$~\citep{Zuo2018}. The H atom accretion flux in these translucent clouds is found to be 
$f2=1.3\times 10^{-8}$ ML s$^{-1}$.

Diffusion barrier and desorption energy are the key parameters that determine the rates of 
surface species diffusion and desorption respectively. The desorption energy of H$_2$ on 
amorphous silicate surfaces, $E_{D1_{H_2}}$, 
was found to follow a continuous distribution based on experimental analysis~\citep{He2011}. On the other hand, 
\citet{He2011} found the value of $E_{D1_{H_2}}$ can be represented by three discrete values, 35, 53 and 75 mev.
The lowest value, 35 mev is about $40\%$ lower than the intermediate value, 53 mev while 
the highest value, 75 mev is about $40\%$ higher than the intermediate value. 
\citet{Hama2012} experimentally studied the diffusion barrier of H atom, $E_{b2_{H}}$ on ASW surfaces 
and found that $E_{b2_{H}}$ should be represented by at least three discrete values, $<18$ mev, 22 mev and $>30$ mev.
The lowest value, $<18$ mev is more than $18\%$ lower than the intermediate value, 22 mev while the 
the highest value, $>30$ mev, is about $40\%$ higher than the intermediate value.

Since the aforementioned experimental findings suggest that three categories of binding sites 
can represent all the various binding sites on surfaces,
we assume the desorption energies of a surface species on dust grain without H$_2$ existence  
always take three discrete values in models.
The lowest value is $40\%$ lower than the intermediate one while the highest value is $40\%$ higher than 
the intermediate one. 
Hereafter, following nomenclature by \citet{Hama2012}, the binding sites with the lowest, intermediate 
and highest desorption energy are called the shallow, middle and deep potential sites.
For simplicity, the number of shallow, middle and deep potential sites are assumed to be equal
on grain surfaces in our models. So, the desorption energy of the middle potential sites is 
the average desorption energy of all binding sites on grain surfaces. 
The three discrete desorption energies are taken from literature if they are available. For instance,
we adopt the three discrete H$_2$ desorption energies on silicate surfaces, which were suggested by \citet{He2011} 
in our models. On the other hand, if only a single desorption energy for a surface species is 
available in literature, we assume that single value to be the desorption energy in the middle potential 
sites.

In a recent review~\citep{Wakelam2017a}, the suggested desorption energies of H atoms
on silicate and ASW surfaces are 44 and 50 mev respectively. These two values are adopted
for the H atom desorption energies in the middle potential sites for the silicate and 
water ice surfaces respectively. The desorption energy of H$_2$ on water ice surfaces varies much in literature.
We adopt the commonly used value of $E_{D2_{H_2}}=440$ K~\citep{Cuppen2007}.
Moreover, in order to study the effect of  $E_{D2_{H_2}}$ on recombination, we also adopt a much larger value, 
$E_{D2_{H_2}}^{'}=800$ K~\citep{Wakelam2017b}.
These two values are assigned to the desorption energy of H$_2$ in the middle potential sites on water ice surfaces 
in translucent cloud models. 

The suggested diffusion barrier of H atoms on the silicate surfaces is 35 mev (406 K)~\citep{Wakelam2017a}. 
So, the ratio of the H atom diffusion barrier to desorption energy is $R=35/44\sim 0.8$. 
We assume R is fixed to be 0.8 for the shallow, middle and deep potential sites on silicate surfaces.
Moreover, because the diffusion
barrier of H$_2$, $E_{b1_{H_2}}$, on the silicate surfaces is not available in literature, 
we assume $E_{b1_{H_2}} = 0.8 E_{D1_{H_2}}$.
Therefore, R is 0.8 for all species on silicate surfaces in our models.

\citet{Hama2012} found that diffusion barrier of H atoms in the middle potential sites on ASW surfaces is 22 mev (255 K). 
So, on the water ice middle potential sites, R is $22/50\sim 0.44$. 
We assume R=0.44 for both H and H$_2$ on all water ice binding sites in our models. 
The H atom diffusion barrier on the shallow water ice sites is $40\%$ lower than that in the middle potential sites.
The adopted energy value is 13.2 mev (153 K), which is lower than its upper bound suggested by \citet{Hama2012}. 
The adopted H atom diffusion barrier on the deep potential sites is 30.8 mev (357 K), which is larger than its lower 
bound suggested by \citet{Hama2012}.
The diffusion barrier of H$_2$ on water ice middle potential sites is, 
$E_{b2_{H_2}} = 0.44 E_{D2_{H_2}}$ or $E_{b2_{H_2}}^{'} = 0.44 E_{D2_{H_2}}^{'}$. 

Table \ref{table1} summarizes the desorption energies and diffusion barriers on the 
shallow, middle and deep potential sites, which are not occupied by H$_2$, in our models.

\begin{table*}
	\caption{Surface species desorption energies and diffusion barriers on silicate and water ice surfaces}
\begin{tabular}{llllll}
\hline
  &  H$_2$ on silicates  &   H$_2$ on water ice$^1$   &  H$_2$ on water ice$^1$ & H on silicates &  H on water ice \\ \hline
desorption energy (K) &  (406, 615, 870) &   (264, 440, 616) & (480, 800, 1120) & (306, 510, 715)     &  (348, 580, 812)    \\ \hline
	diffusion barrier (K) &  (325, 492, 696) & (116, 194, 271) & (211, 352, 493) & (244, 406, 568)         &  (153, 255, 357)         \\
\hline
\label{table1}
\end{tabular}
\medskip{\protect\\
Notes.\protect\\
All energy values are in K. 1 mev = 11.6 K. (a, b, c) are desorption energies or desorption barriers for the 
	shallow, middle and deep potential sites respectively. 
 $^1$ These two sets of energy values are used in different models, which will be explained in detail later in this section.
\protect\\
}
\end{table*}

\begin{figure*}
\includegraphics[scale=0.4]{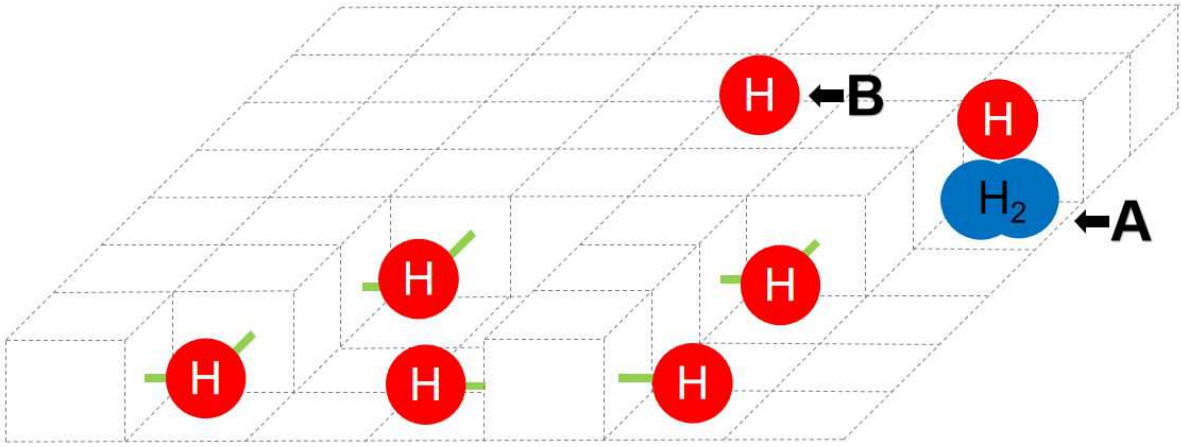}
\caption{Lateral bonds formed by H atoms and their neighbors in different sites. In sites other than A and B, H atoms 
	can form lateral bonds with their horizontal neighbors. Site A is occupied by H$_2$, so the H atom in this site
	cannot form lateral bonds with its horizontal neighbors. Site B is a binding site in which a H atom cannot 
	see horizontal neighbors, thus, no lateral bonds can be formed between the H atom and its horizontal neighbors.
}
\label{Fig1}
\end{figure*}

The desorption energies of H$_2$ and H on H$_2$ substrate are much lower than these on silicate or ASW surfaces. 
At 10 K, the desorption energy of H on H$_2$ substrate 
was calculated to be $E_{D_{HH_2}}=45$ K~\citep{Pierre1985,Vidali1991,Cuppen2007}. 
\citet{Cuppen2007} estimated that the desorption energy of H$_2$ on H$_2$ substrate is, $E_{D_{H_2H_2}}=23$ K.
Recently, by performing quantum chemical calculations, 
\citet{Das2021} found that $E_{D_{H_2H_2}}$ is between 67 K and 79 K while $E_{D_{HH_2}}$ is between 23 K and 25 K. 
Moreover, if a site has been occupied by H$_2$, a H atom in this site may not be able to 
see horizontal neighbors, thus cannot form lateral bonds with
its horizontal neighbors. So, the desorption energy of the surface species is further reduced 
when the existence of H$_2$ is considered. Fig. \ref{Fig1} illustrates lateral bonds formed between H atoms and their
horizontal neighbors in different binding sites.

Not only H$_2$ molecules but also H atoms can enhance the desorption rates of H$_2$. \citet{Cuppen2007} estimated that the desorption energy of H$_2$
on H substrate is, $E_{D_{H_2H}}=30$ K, which is also significantly lower than that on silicate surfaces. In all our models,
if H$_2$ molecules resides in sites occupied by H atoms, the desorption energy of H$_2$ on these sites become 30 K. 

Models M01-M06 are used 
to study H$_2$ formation on the olivine grains in diffuse clouds. Model M01 is a reference model that is similar 
to earlier models~\citep{Chang2005,Cuppen2005}. Gas-phase H$_2$ are allowed to accrete on grain surfaces in models M02-M06, but not in model M01.
When a exothermic chemical reaction occurs on grain surfaces, products may desorb immediately, which is known 
as chemical desorption~\citep{Garrod2007}.
The chemical desorption coefficient, $\mu$, measures the percentage of products that desorb immediately after a reaction fires.  
In the reference model M01, $\mu$ is set to be 1.
In model M02, the desorption energies of H$_2$ and H on H$_2$ substrate are 23 K and 45 K respectively. 
Moreover we assume a surface species loses all lateral bonds with its horizontal 
neighbors if the surface species is in a site already occupied by H$_2$.
The chemical desorption coefficient is set to be 1 in model M02.
To study the effect of $\mu$ value on H$_2$ formation efficiency, we simulate model M03, which is similar to model M02, but
$\mu$ is set to be 0. 

Because both $E_{D_{H_2H_2}}$ and $E_{D_{HH_2}}$ are not well known yet,
we simulate models M04 and M05 to study how the value of $E_{D_{H_2H_2}}$ and $E_{D_{HH_2}}$ may affect 
the H$_2$ formation efficiency. 
These two models are similar to model M02.
The difference between models M02 and M04 is that the desorption energies of H and H$_2$ on H$_2$ 
substrate are set to be 24 K and 73 K respectively in model M04 based on the energy values suggested by \citet{Das2021}. 
Since the residence time of H atoms on dust grains increases with $E_{D_{HH_2}}$, 
H$_2$ formation efficiency at higher temperatures should increase as $E_{D_{HH_2}}$ becomes larger. 
We would like to set the value of $E_{D_{HH_2}}$ to be as high as possible in model M05.
However, to the best of our knowledge, the upper bound of $E_{D_{HH_2}}$ is not available in literature. 
We argue that the desorption energy of H atoms on the silicate shallow potential sites may be 
a upper bound for $E_{D_{HH_2}}$ for the following reasons.
Hydrogen atoms in sites occupied by H$_2$ can hardly form lateral bonds with 
their horizontal neighbors in sites occupied by H$_2$ as shown in Fig.~\ref{Fig1}. 
So, even if the vertical bond formed between H atoms and H$_2$ substrate is as strong as that between 
H atoms and silicate surfaces, $E_{D_{HH_2}}$ cannot be higher than the 
desorption energy of H atoms on the silicate shallow potential sites.
Similarly, the desorption energy of H$_2$ on the silicate shallow potential sites may be a upper bound for $E_{D_{H_2H_2}}$.
In model M05, both $E_{D_{H_2H_2}}$ and $E_{D_{HH_2}}$ are set to be their upper bounds.

In models M02-M05, deep H atom potential sites must also be deep H$_2$ potential sites. This assumption
is based on our current understanding about how stronger binding sites are formed~\citep{Cuppen2005}, which might be limited.
Therefore, we also simulated model M06, in which deep potential sites for H atoms cannot be deep potential 
sites for H$_2$ molecules. The deep H atom potential site can be a middle or shallow potential site 
for H$_2$ molecules with equal probability. Similarly, it is equally likely that 
a deep H$_2$ molecule potential sites is a middle or shallow potential site for H atoms.
In models M02-M06, the ratio of a surface species diffusion barrier to its desorption energy on sites occupied by H$_2$, 
R$_{H_2}$ is fixed to be 0.8. Moreover, the ratio of H$_2$ diffusion barrier to its desorption 
energy on sites occupied by H atoms, R$_{H}$ is also 0.8. 

Models M11-M15 are used to study H$_2$ formation on dust grains in translucent clouds. In all these models, $\mu$ is set to be 1. 
The model M11 is a reference model in which H$_2$ cannot accrete while H$_2$ are allowed to accrete in models M12-M15.
The desorption energies of H and H$_2$ on sites occupied by H$_2$ are 45 K and 23 K respectively in models M12 and M13. 
In model M12, R$_{H_2}$ and R$_{H}$ are fixed to be 0.44, which is the same as the value of R on water ice binding sites without H$_2$. 
These two ratios are 0.8 in model M13 to study how R$_{H_2}$ and R$_{H}$ may affect the recombination efficiency.
The H$_2$ desorption energy on water ice is 440 K in models M11-M13.
To study how the H$_2$ desorption energy on water ice may affect the recombination efficiency on grain surfaces,
we simulated models M14 and M15.
The H$_2$ desorption energy is 800 K in these two models.
In model M14, R$_{H_2}$ and R$_{H}$ are 0.44 while in model M15, these two ratios are 0.8. 

Our purpose is to make the reference models to be as close to the earlier models~\citep{Chang2005, Cuppen2007} 
as possible for comparison, 
so, some of the assumptions in the earlier models
are kept in our models. The H atom quantum tunneling effect was not considered in the earlier models because 
H atoms diffuse mainly by thermal hopping on silicate or ASW surfaces, so this effect is also ignored in our models. 
The rate of hopping is $\nu \exp(-E_b/T)$, 
where $E_b$ is the diffusion barrier, T is the dust grain temperature,
the pre-exponential factor $\nu$ is $10^{12}$ s$^{-1}$. 
The rate of desorption is  $\nu \exp(-E_D/T)$ where $E_D$ is the desorption energy.
When H atoms land on empty sites or sites occupied by H$_2$, the sticking coefficient is 1. 
However, H atoms are not allowed to accrete on sites already occupied by H atoms. 
On the other hand, we assume the H$_2$ sticking coefficient is always 1 regardless of the landing sites.

Table \ref{table2} summarizes models in this work.

\begin{table*}
\caption{Summary of models}
\begin{tabular}{llllllllll}
\hline
	& astronomical  & H$_2$       & R  &  R$_{H_2}$=R$_{H}$ & H$_2$ desorption  & H atom desorption   & H$_2$ desorption & $\mu$ & same?$^1$ \\ 
	& sources        & accretion?  &    &            & energy on H$_2$   & energy on H$_2$     & energy on middle potential &       &      \\
	&	        &             &    &            & substrates (K)    & substrates (K)     & water ice sites (K)   &       & \\ \hline
M01     &  diffuse      & no          & 0.8 & N.A.      & N.A.              & N.A.                & N.A.               & 1      & yes  \\  
	&  clouds       &             &     &           &                   &                     &                    &        &    \\  \hline
M02     &  diffuse      & yes          & 0.8 & 0.8      & 23              & 45                & N.A.               & 1      & yes  \\  
	&  clouds       &             &     &           &                   &                     &                    &        &    \\  \hline
M03     &  diffuse      & yes          & 0.8 & 0.8      & 23              & 45                & N.A.               & 0      & yes  \\  
	&  clouds       &             &     &           &                   &                     &                    &        &    \\  \hline
M04     &  diffuse      & yes         & 0.8 & 0.8      & 73              & 24                & N.A.               & 1      & yes  \\  
	&  clouds       &             &     &           &                   &                     &                    &        &    \\  \hline
M05     &  diffuse      & yes          & 0.8 & 0.8      & 406               & 306                & N.A.               & 1      & yes  \\  
	&  clouds       &             &     &           &                   &                     &                    &        &    \\  \hline
M06     &  diffuse      & yes          & 0.8 & 0.8      & 23              & 45                & N.A.               & 1      & no  \\  
	&  clouds       &             &     &           &                   &                     &                    &        &    \\  \hline
M11     &  translucent  & no          & 0.44 & N.A.      & N.A.              & N.A.                & N.A.               & 1      & yes  \\  
	&  clouds       &             &     &           &                   &                     &                    &        &    \\  \hline
M12     &  translucent  & yes          & 0.44 & 0.44      & 23              & 45                & 440              & 1      & yes  \\  
	&  clouds       &             &     &           &                   &                     &                    &        &    \\  \hline
M13     &  translucent  & yes         & 0.44 & 0.8      & 23              & 45                & 440               & 1      & yes  \\  
	&  clouds       &             &     &           &                   &                     &                    &        &    \\  \hline
M14     &  translucent  & yes          & 0.44 & 0.44      & 23              & 45                & 800               & 1      & yes  \\  
	&  clouds       &             &     &           &                   &                     &                    &        &    \\  \hline
M15     &  translucent  & yes         & 0.44 & 0.8      & 23              & 45                & 800               & 1      & yes  \\  
	&  clouds       &             &     &           &                   &                     &                    &        &    \\  \hline

\label{table2}
\end{tabular}
\medskip{\protect\\
Notes.\protect\\
$^1$ same means deep H atom potential sites are also deep H$_2$ potential sites. N.A. means the parameter is not applicable to the model.
\protect\\
}
\end{table*}

\section{Numerical Method}\label{sec:method}
In this work, we use the microscopic Monte Carlo (MMC) method to perform model simulations. 
The MMC method is a rigorous approach that can treat the finite size effect.
In this section, we only briefly introduce this method and refer to \citet{Chang2005} and \citet{Chang2012} for details.

To perform MMC simulations, binding sites on grains are put on 
a $L\times L$ square lattice, where L is the square root of the total number of sites on the grain.
Previous studies show that L=100 is already large enough to eliminate finite size effect~\citep{Chang2005}, so
in this work, L is fixed to be 100.
Each site has four nearest neighbor sites on the square lattice.
Gas-phase species can accrete and become surface species on the lattice.
We keep track of the position of each surface species, which hops from one site to one of 
its nearest neighbor. A chemical reaction occurs when two reactive species are in the same 
site on the lattice. Other than hopping, surface species can also desorb. To decide whether
a surface species hops or desorbs, we generate a random number, X, that is uniformly distributed
between 0 and 1. Assuming the hopping and desorption rates of the surface species 
are $b_1$ and $b_2$ respectively, the species hops if $X<\frac{b_1}{b_1+b_2}$; otherwise, it desorbs.
There are three types of events on the lattice, accretion, hopping and desorption.
We keep track of the absolute time when each event occurs. The smallest absolute time is  
sorted out and the event that occurs at this time is executed. 

The absolute time when each event happens are calculated as the follows. 
In MMC simulations, after an event occurs, the same event will occur again 
after a ``waiting time'', $\tau = - ln(Y)/k$, where Y is another random number 
uniformly distributed within (0, 1) while k is the rate of the event.  
So, assuming a surface species hops from one site to another at time t, 
the absolute time it hops again or desorbs 
is calculated as,  $t^{'} = t - ln(Y)/(b_1+b_2)$.
Similarly, the absolute time when an accretion event of a gas-phase species occurs 
is, $t_{acc} = t_{acc}^{'} -  ln(X^{'})/k_{acc}$, where $X^{'}$ is a random number uniformly 
distributed within (0, 1), $t_{acc}^{'}$ is the absolute time when last accretion event occurs 
(it is 0 in the beginning of simulations)
while $k_{acc}$ is the accretion rate of the species.

In the beginning of model simulations, the population of surface H and H$_2$ fluctuates
but also increases. After some time, the steady state condition is reached when 
both H and H$_2$ population fluctuates around the converged value. 
After steady states have been reached, we choose a time period $\tau^{,}$ to calculate 
the recombination efficiency, $\eta$, which measures how efficiently H$_2$ are formed on grain surfaces.
We count the number of H$_2$ formed, $N_{H_2}$ and the number of H atom accretion events, $N_H$
during $\tau^{,}$. The recombination efficiency is calculated as,
$\eta = \frac{2N_{H_2}}{N_H}$. We set $\tau^{,}$ large enough so that the value of $\eta$ converges.

\section{Results}\label{sec:res}

\subsection{Diffuse clouds}

The H$_2$ molecules were observed to be formed with rate coefficient 3-4$\times 10^{-17}$ cm$^3$ s$^{-1}$ 
in the diffuse interstellar medium~\citep{Gry2002,Wakelam2017a}. The H$_2$ formation rate coefficient 
3$\times 10^{-17}$ cm$^3$ s$^{-1}$ corresponds to a recombination efficiency of 0.5 on dust grains~\citep{Gry2002,Hollenbach1971a}. 
So, a recombination efficiency of 0.5 is the minimum to explain observations toward diffuse clouds.
Hereafter, efficient H$_2$ formation means $\eta\geq0.5 $ in this work. 

\begin{figure*}
\includegraphics[scale=0.5]{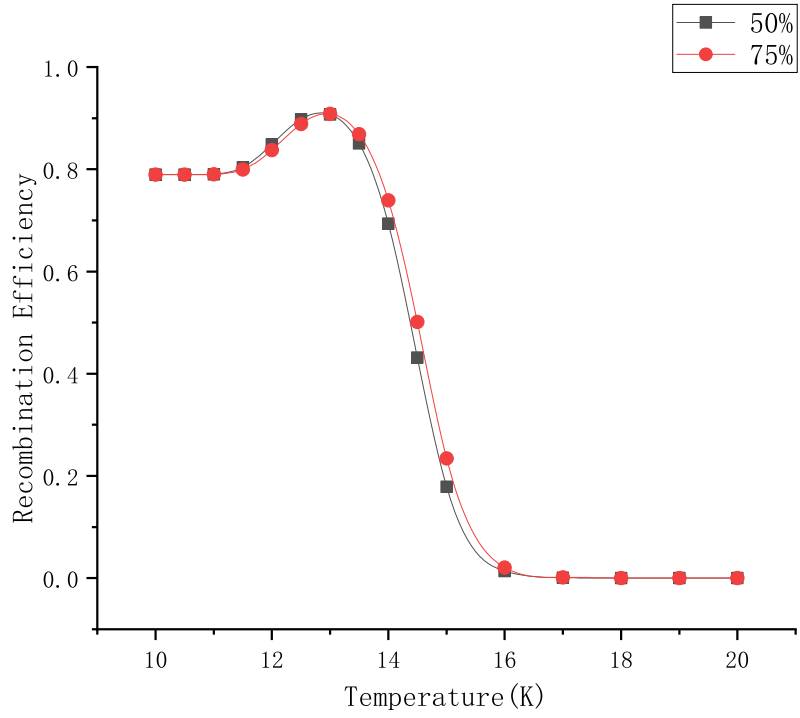}
\caption{The calculated recombination efficiency, $\eta$ as a function of grain temperature, T using model M01.
Circles and squares represent $\theta=0.75 $ and 0.5 respectively. 
}
\label{fig2}
\end{figure*}

Fig. \ref{fig2} shows the calculated recombination efficiency, $\eta$ as a function of grain surface temperature, T in the reference model M01.
Similar to previous models~\citep{Chang2005,Cuppen2007},
H atoms can be trapped in the stronger binding sites, 
so H$_2$ can be efficiently formed on grains over a temperature range between 10 K and 14 K in this model.    
At temperatures between 10 and 13 K, $\eta$ increases with T because 
surface H atom population decreases at higher temperatures, so more H atoms are able to accrete on grain surface instead of being rejected
to the gas-phase. This phenomenon was also reported by previous studies~\citep{Chang2005}. At temperatures above 13 K, $\eta$ decreases as T increases
because of the enhanced desorption rate of H atoms.
In model M01, we assume three different values of $\theta=n_{H}/(2n_{H_2}+n_H)$, where $n_{H_2}$ and $n_H$ are the number
densities of gas-phase H and H$_2$ respectively. 
We can see that the value of $\theta$ does not affect $\eta$ much in the model M01. 

\begin{figure*}
\includegraphics[scale=0.5]{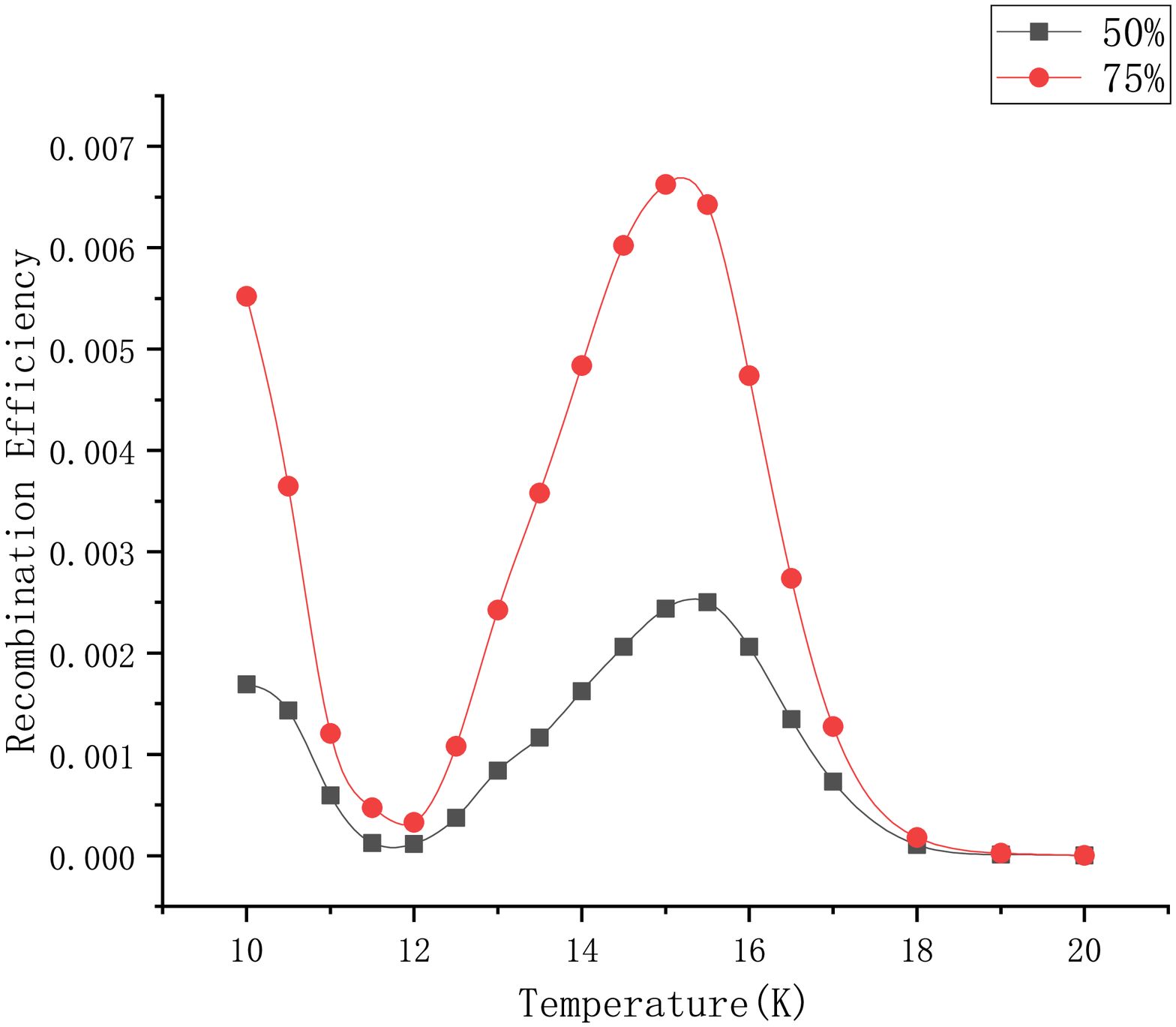}
\caption{The calculated recombination efficiency, $\eta$ as a function of grain temperature, T using model M02.
Circles and squares represent $\theta=0.75 $ and 0.5 respectively. 
}
\label{fig3}
\end{figure*}

The existence of H$_2$ on grain surface in model M02 can significantly reduce the recombination efficiency as show in Fig. \ref{fig3}.  
The decrease is especially severe at around 12 K, where the recombination efficiency drops to nearly 0.
Efficient H$_2$ formation is not possible for all T $\geq 10 $ K in model M02. 
In this model, the desorption rate of H atoms increases due to surface H$_2$ although H atoms cannot react with H$_2$,
so $\eta$ in model M02 is more than one order of magnitude lower than that in model M01.
On the other hand, we can see that $\eta$ as a function of T in model M02 is more complex than in model M01.
The reason is the following. Higher T can increase H$_2$ desorption rates, so the population of 
H$_2$ decreases as the T increases. So, H atoms are less likely to encounter H$_2$ and then desorb
at higher T. Moreover, there are more stronger
sites available on grain surfaces for H atoms to occupy and then be trapped as the H$_2$ population decreases.
Therefore, $\eta$ should increase if we only consider the decrease of surface H$_2$ population at higher T.
On the other hand, higher T can also increase the thermal desorption rate of H atoms, which
decreases $\eta$. Fig. \ref{fig3} shows that the net effect of increasing T on $\eta$ is dependent on T. 
At grain temperatures lower than 12 K, $\eta$ decreases as T increases. 
However, at grain temperatures between 12 K and 15 K, $\eta$ increases with T. 
As T increases beyond 15 K,  $\eta$ decreases as T increases.
Finally, we assumed two values of $\theta$, 0.75 and 0.5 in the model M02.
Fig. \ref{fig2} shows that the recombination efficiency becomes larger as  $\theta$ becomes larger in the model M01. 
The reason is that the accretion rate of H$_2$ is smaller when $\theta$ is larger, 
so the population of surface H$_2$ decreases, which helps H atoms to be trapped on grain surfaces.

\begin{figure*}
\includegraphics[scale=0.5]{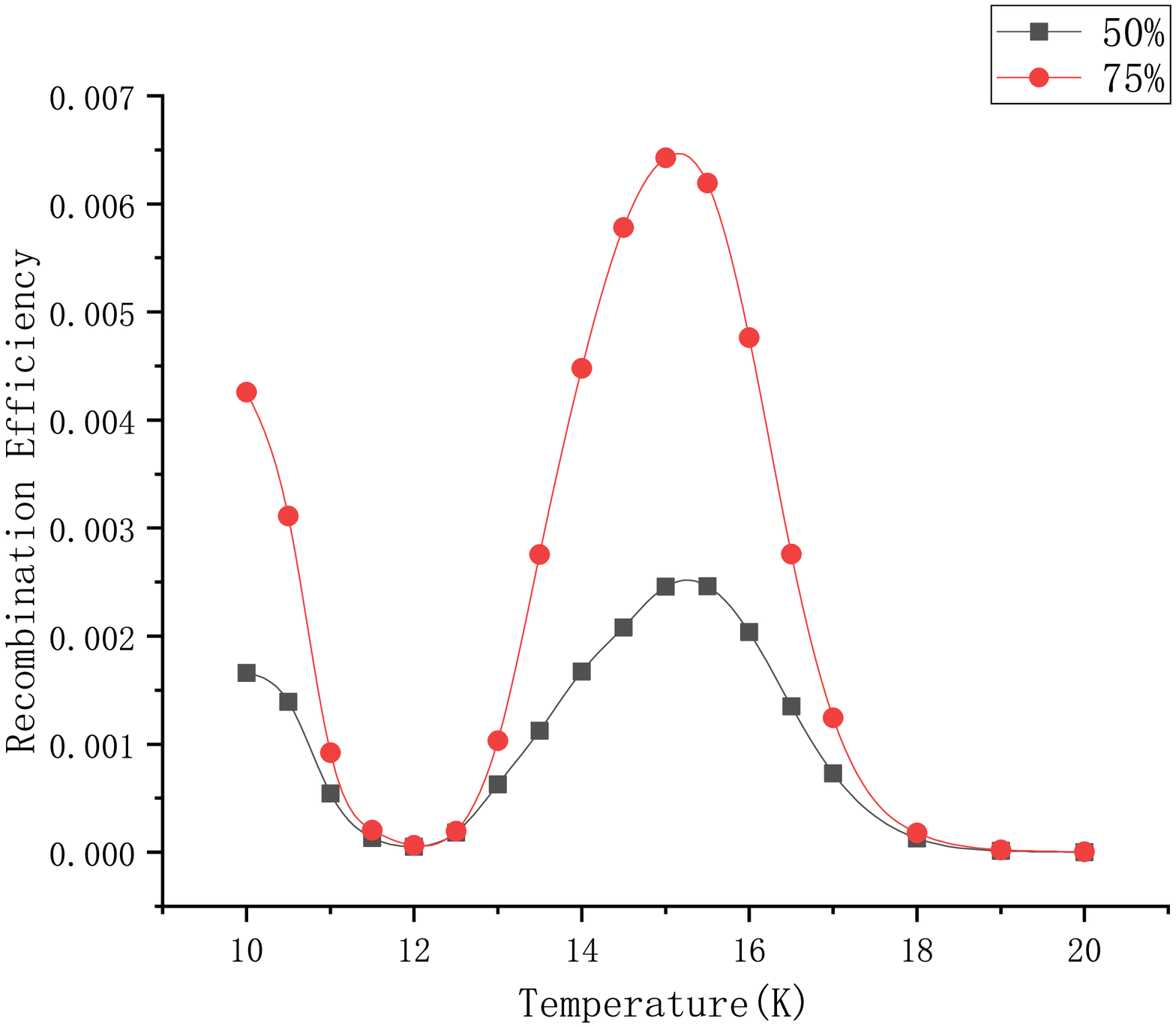}
\caption{The calculated recombination efficiency, $\eta$ as a function of grain temperature, T using model M03.
Circles and squares represent $\theta=0.75 $ and 0.5 respectively. 
}
\label{fig4}
\end{figure*}

Fig. \ref{fig4} shows the recombination efficiency predicted by model M03 as a function of T. 
By comparing models 
M02 and M03 results, we see that the chemical desorption coefficient, $\mu$ can hardly 
affect $\eta$ for both  $\theta=0.5$ and 0.75.
The reason is that the major source of surface H$_2$ is the accretion of H$_2$ from gas phase. Therefore, 
we assume $\mu$ to be 1 for all models other than M03.

\begin{figure*}
\includegraphics[scale=0.5]{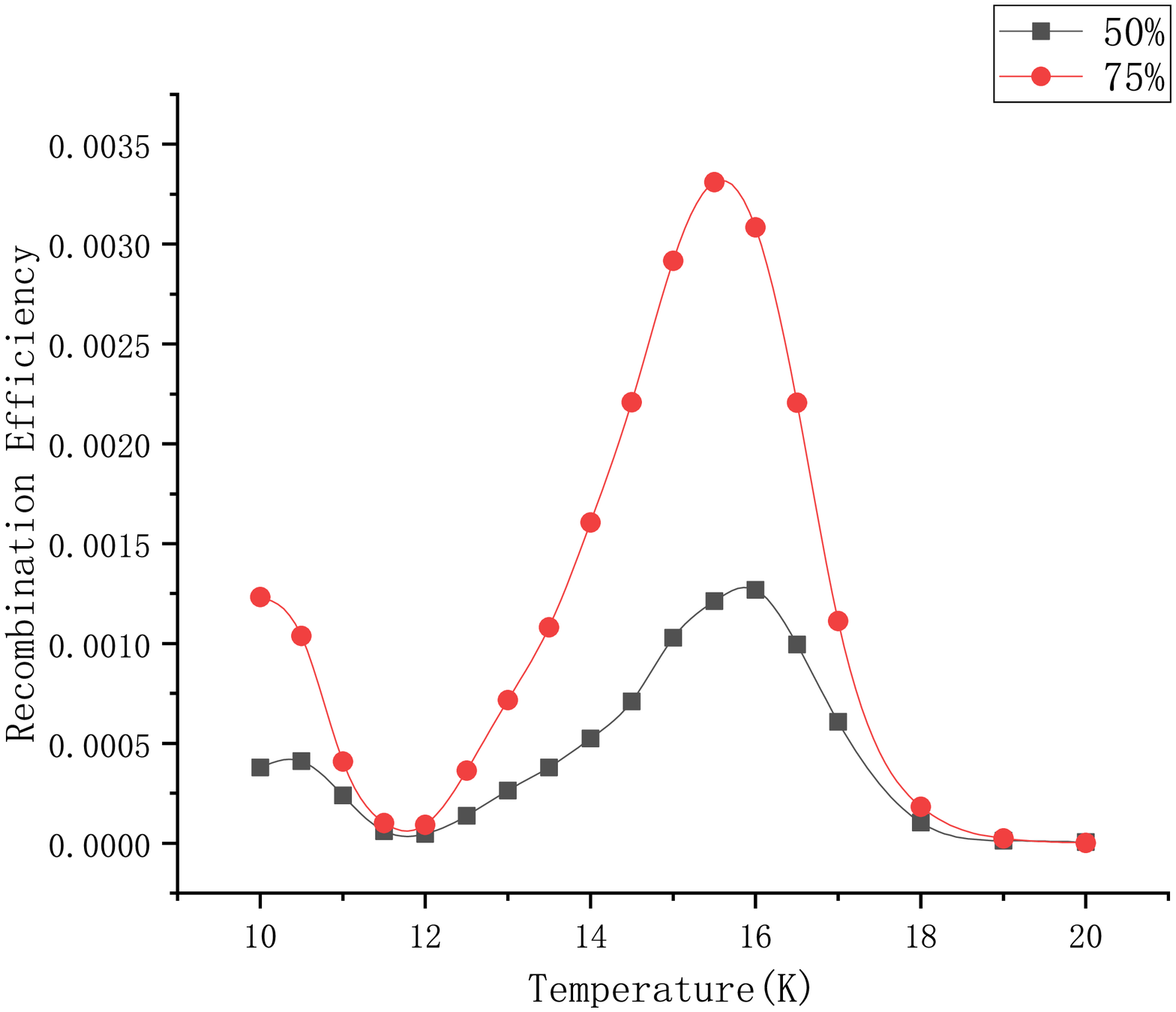}
\caption{The calculated recombination efficiency, $\eta$ as a function of grain temperature, T using model M04.
Circles and squares represent $\theta=0.75 $ and 0.5 respectively. 
}
\label{fig5}
\end{figure*}

Fig. \ref{fig5} shows the recombination efficiency predicted by model M04 as a function of grain temperature.
The model M04 adopts the more up-to-date desorption energies of H atoms and H$_2$ molecules on H$_2$ substrates.  
Recombination efficiency in model M04 is typically lower than model M02 for both $\theta=0.75$ and 0.5.
For instance, at 15 K, $\eta$ in model M04 is about half of that in model M02.
The reason is that the desorption energy of H atoms on H$_2$, $E_{D_{HH_2}}$ in the model M04 
is lower than that in model M02, so H atoms desorb more quickly when they encounter H$_2$. 
Moreover, because of the higher desorption energy of H$_2$ molecules on H$_2$ in model M04, 
the population of H$_2$ in the model M04 is larger than the model M02, so H atoms are more likely to encounter
H$_2$ and then desorb in the model M04.

\begin{figure*}
\includegraphics[scale=0.5]{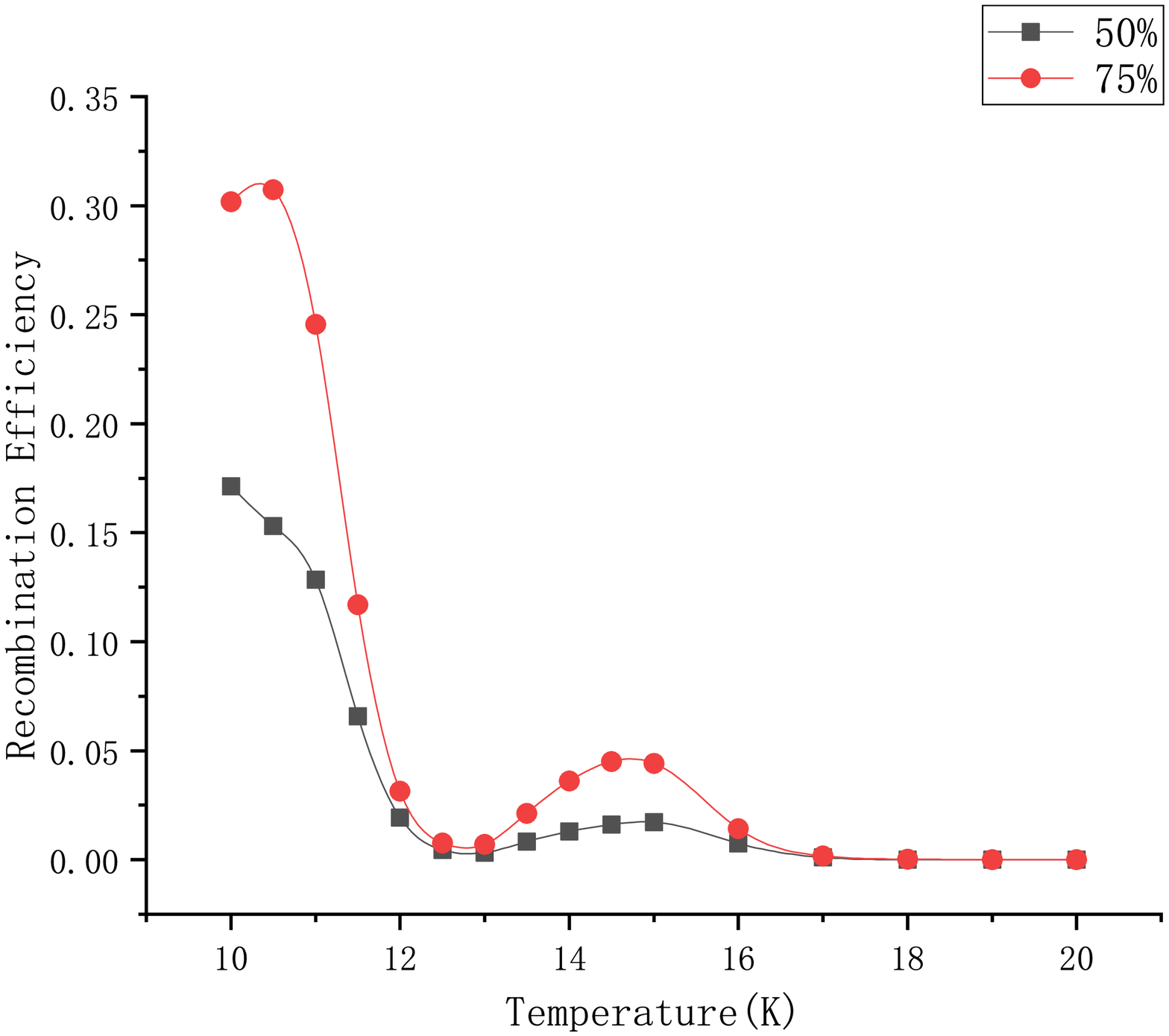}
\caption{The calculated recombination efficiency, $\eta$ as a function of grain temperature, T using model M05.
Circles and squares represent $\theta=0.75 $ and 0.5 respectively. 
}
\label{fig6}
\end{figure*}

Fig. \ref{fig6} shows the recombination efficiency predicted by model M05 as a function of T.  
Compared with models M02-M04 results, the recombination efficiency predicted by model M05 increased significantly
because of the much larger H atom desorption energy on sites occupied by H$_2$. 
However, efficient H$_2$ formation is still not possible on dust grains if T is above 10 K 
for both $\theta=0.5$ and 0.75. 

\begin{figure*}
\includegraphics[scale=0.5]{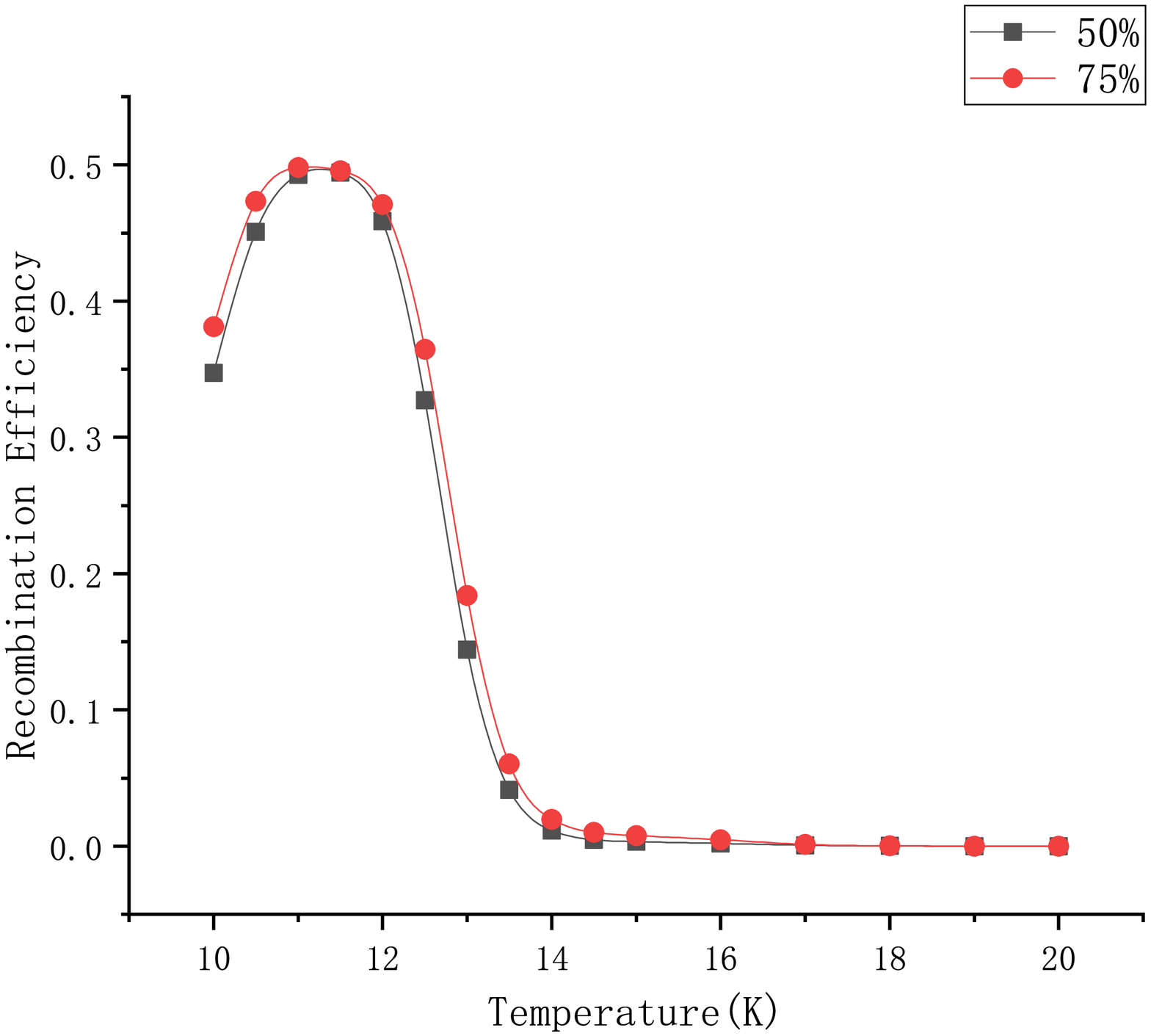}
\caption{The calculated recombination efficiency, $\eta$ as a function of grain temperature, T using model M06.
Circles and squares represent $\theta=0.75 $ and 0.5 respectively. 
}
\label{fig7}
\end{figure*}

Fig. \ref{fig7} shows the recombination efficiency predicted by model M06 as a function of T.
Hydrogen atom can still efficiently desorb when they encounter H$_2$ on grain surfaces,
although there are more deep potential sites available for them to occupy in this model. Therefore,
over all, the recombination efficiency predicted by this model is lower than model M01, but larger than by models M03-M05.
Moreover, at T above 12 K, $\eta$ predicted by model M06 decreases more quickly than by model M02.
So, in model M06, efficient H$_2$ formation occurs only around T=11 K for both $\theta=0.5$ and 0.75. 

\subsection{Translucent clouds}

\begin{figure*}
\includegraphics[scale=0.5]{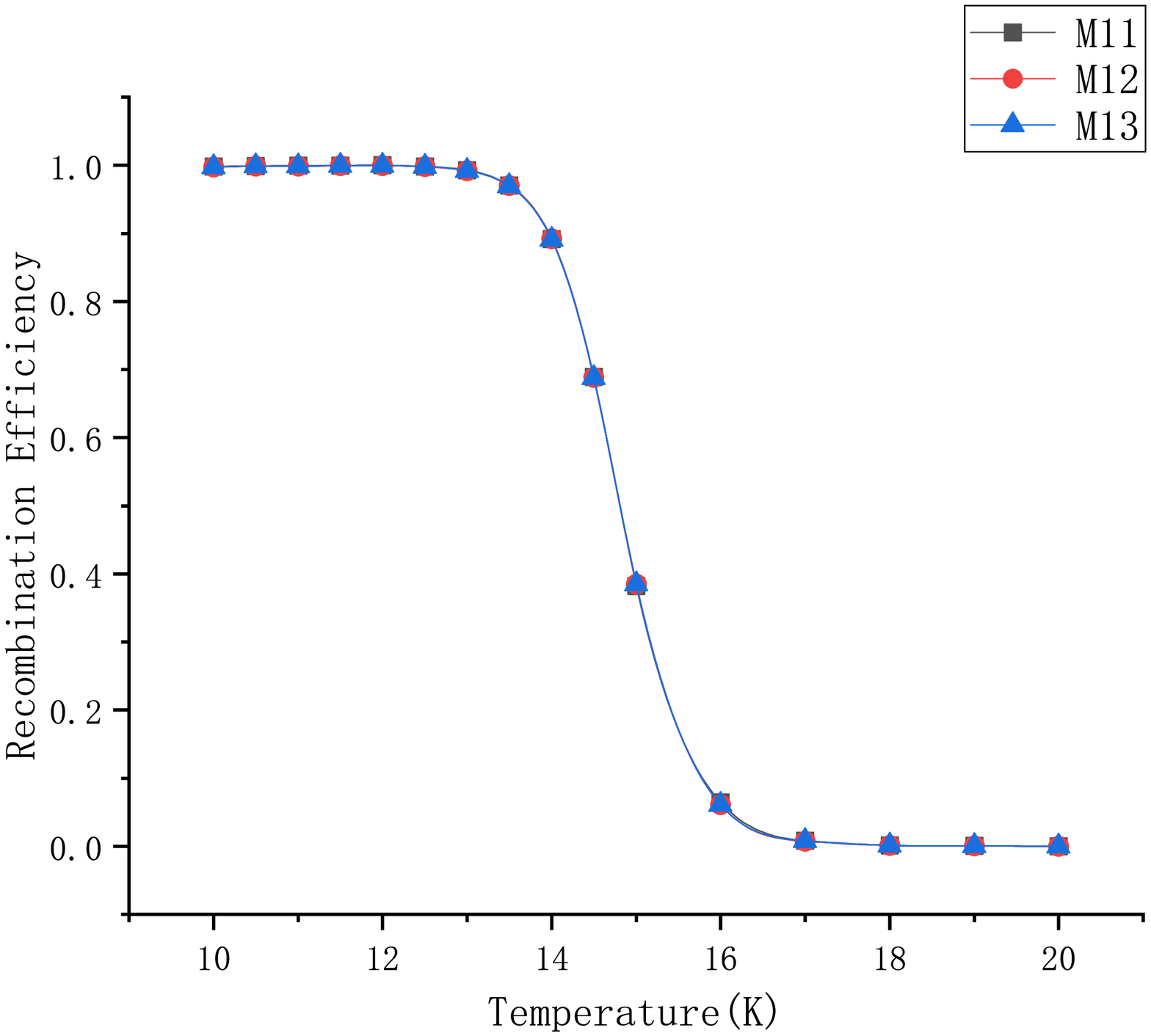}
\caption{The calculated recombination efficiency as a function of grain temperature using models M11-M13. 1-$\theta=$0.75 in all models.
}
\label{fig8}
\end{figure*}

Fig.~\ref{fig8} shows the recombination efficiency predicted by models M11-M13 as a function of grain temperature. 
In the reference model M11, H$_2$ can be efficiently synthesized until T is close to 15 K. 
The recombination efficiency predicted by model M12 is almost the same as that by the reference model M11. 
The reason is the following.
First, the diffusion barriers of H$_2$ on water ice sites are 
low enough so that H$_2$ cannot be trapped by the deep potential sites while H atoms can be trapped in the deep potential
sites in model M12.
Secondly, when H$_2$ molecules diffuse and encounter H atoms, their desorption energies significantly drops, so 
H$_2$ can desorb quickly in this model. So, there are only a few H$_2$ molecules on the lattice.
Therefore, H atoms can hardly desorb by encountering H$_2$ on the lattice, thus, $\eta$ predicted by models 
M11 and M12 is almost identical.
On the other hand, we can see that the recombination efficiency can hardly be affected by the variation of R$_{H_2}$ and R$_{H}$ 
values in models M13.  
The small H$_2$ population on the lattice makes it rare for H atoms to have opportunity to encounter H$_2$ molecules.
So, the value of R$_{H_2}$ or R$_{H}$ is not important in model M14. 

\begin{figure*}
\includegraphics[scale=0.5]{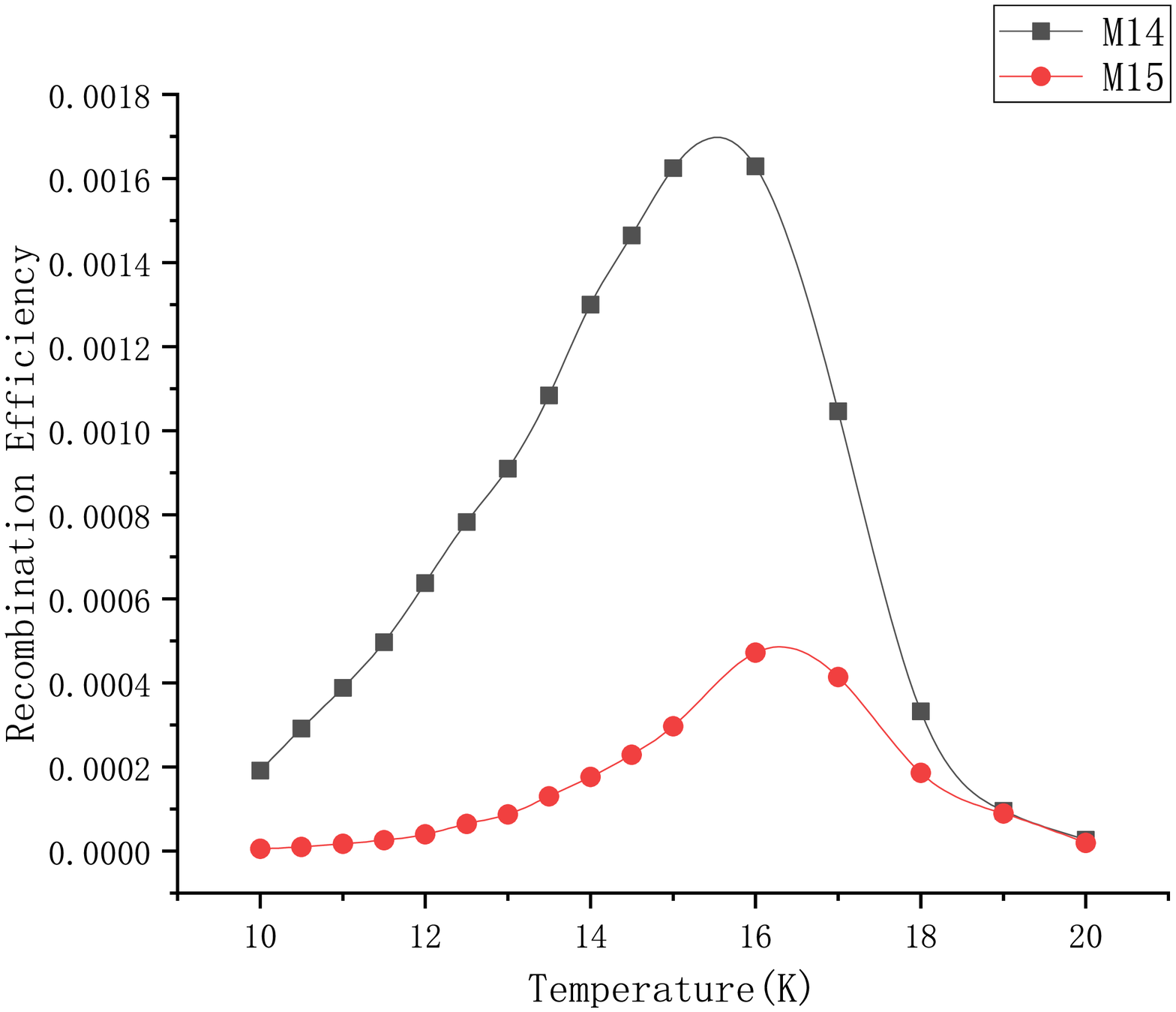}
\caption{The calculated recombination efficiency as a function of grain temperature using models M14 and M15. 1-$\theta=$0.75 in all models
}
\label{fig9}
\end{figure*}

Fig.~\ref{fig9} shows the recombination efficiency predicted by models M14-M15 as a function of T.
The diffusion barriers on water ice sites adopted by these models are almost a factor of two larger than those in models M11-M13.   
The enhanced diffusion barriers enable H$_2$ to be trapped by the deep potential sites, which significantly increases
the H and H$_2$ interaction on grain surfaces. The recombination efficiency predicted by both models 
is less than 0.002 and is much lower than that by models M11-M13.
Molecular hydrogen cannot be efficiently synthesized at temperatures above 10 K 
in both models. 

The recombination efficiency predicted by model M15 is lower than that by model M14. 
The diffusion barrier of H atoms on H$_2$ in model M14 is lower than that in model M15. In the MMC simulations as introduced 
in Section~\ref{sec:method}, if the desorption energy of a species is constant, the probability that a surface species 
desorb decreases as the diffusion barrier of that species decreases. Therefore, H atoms in sites occupied by H$_2$ 
are more likely to diffuse out of these sites instead of desorb in model M14 than in model M15. So model M14 predicts larger
recombination efficiency.

\section{Summary and conclusions} \label{sec:sum}
Although H$_2$ molecules may account for a large fraction of gas phase species in the diffuse and translucent clouds, 
their existence on grain surfaces has been ignored by almost all previous H$_2$ formation models.
To the best of our knowledge, this work is the first time that the detailed MMC method has been used to 
investigate how the existence of H$_2$ may affect the efficiency of molecular 
hydrogen formation on cold interstellar dust grains in diffuse and translucent clouds. 
We found that the impact of the existence of H$_2$ on recombination efficiency mainly depends 
on the diffusion barriers of H$_2$ on grain surfaces. 

In the diffuse cloud models M02-M06, because of the high diffusion barriers on olivine grains, 
H$_2$ molecules can be trapped by the deep potential sites. Site occupied by H$_2$ become much weaker 
binding sites for H in these models. 
So, in models M02-M05, it becomes more difficult for H atoms to become trapped in these sites 
already occupied by H$_2$.
Hydrogen atoms can also efficiently desorb when they encounter H$_2$ molecules in models M02-M05, 
leading to a lower predicted recombination efficiency in models M02-M05 than in the reference 
model M01, which does not allow H$_2$ to exist on grain surfaces. Moreover, models M02-M05 all fail to
predict the minimum recombination efficiency (0.5) to explain observations.

The recombination efficiency increases with the value of $E_{D_{HH_2}}$, which is, however, not well known yet.
The adopted value of $E_{D_{HH_2}}$ in models M02-M04 is much lower than the desorption energy of H atoms 
on silicate surfaces while model M05 adopts a upper bound of $E_{D_{HH_2}}$.
Future more rigorous study might suggest higher values of $E_{D_{HH_2}}$ and $E_{D_{H_2H_2}}$ than these used in models M02-M04.
However, because even model M05 predicts significant reduction of the recombination efficiency,
we argue that even if these higher values of $E_{D_{HH_2}}$ and $E_{D_{H_2H_2}}$ are adopted by 
models M02-M04, these models may still predict lower recombination efficiency than the reference model M01 does.  

In model M06, because H atoms and H$_2$ molecules do not share common stronger binding sites, 
the stronger H atom binding sites can still trap H atoms when H$_2$ molecules are trapped on grain surfaces.  
However, H atoms can still desorb efficiently when they encounter H$_2$.
Model M06 results show that as long as H$_2$ are trapped by the strong binding sites 
and the desorption energy of H atoms on H$_2$ substrate is much lower than that on silicate grain surfaces,
efficient H$_2$ formation ($\eta \geq 0.5$) occurs only when the dust surface temperature is around 11 K.

We used olivine grains as an example in the diffuse cloud models. The diffusion barrier
of H$_2$ on carbon grain surfaces is even larger that on silicate surfaces. So, H$_2$ molecules could
be more tightly trapped on grain surfaces. The tightly trapped H$_2$ molecules should also make H$_2$ formation
on carbon grains less efficient.

In the translucent clouds models M12 and M13, the diffusion barriers of H$_2$ on grains covered by water ice 
are low enough so that H$_2$ molecules diffuse even faster than H atoms and
cannot be trapped by the deep potential sites. Moreover, H$_2$ molecules can efficiently desorb when they encounter 
H atoms on grain surfaces. So H$_2$ population in these two models are very low so that the enhanced H atom desorption 
by encountering H$_2$ molecules is minimal. Therefore, the existence of H$_2$ on grain surfaces can hardly affect 
the recombination efficiency. 
On the other hand, in models M14 and M15, the diffusion barriers of H$_2$ are almost a factor of two larger than these in
models M12 and M13, so H$_2$ can be trapped by the deep potential sites.
In models M14 and M15, efficient H$_2$ formation is not possible when the temperatures are above 10 K.

\begin{figure*}
\includegraphics[scale=0.5]{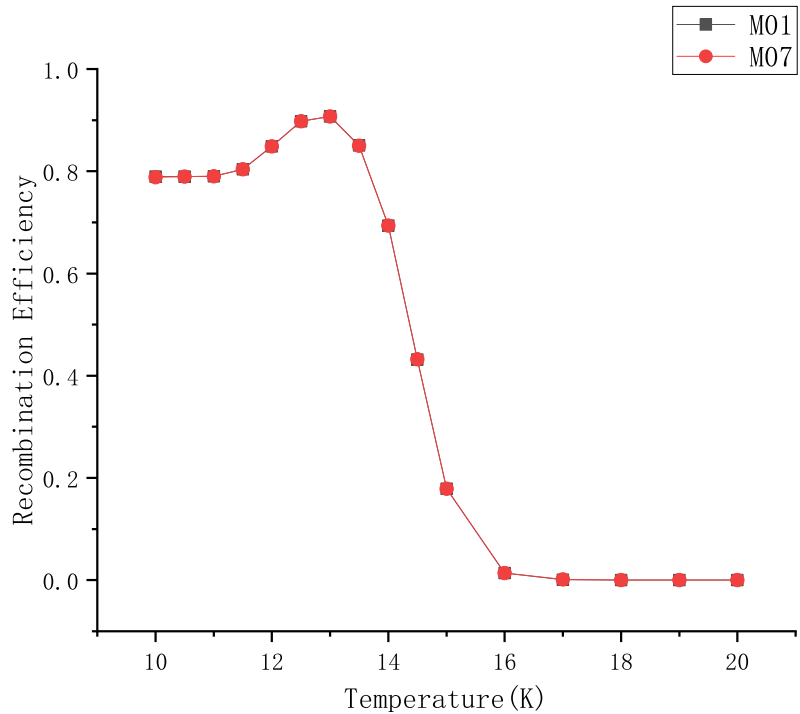}
\caption{The calculated recombination efficiency as a function of grain temperature using models M01 and M07.
$\theta=0.5$ in both models. 
}
\label{fig10}
\end{figure*}

Based on the translucent cloud model results, we can conclude that lower H$_2$ diffusion 
barriers on bare dust grains can enhance the recombination efficiency. 
This conclusion suggests that $\eta$ may also increase 
if lower H$_2$ diffusion barriers is used in diffuse models.
In order to find out how lower H$_2$ diffusion barriers may affect diffuse cloud model results,
we simulate a test diffuse cloud model M07 using reduced H$_2$ diffusion barriers, which  
are $50\%$ of the H$_2$ desorption energies on silicate surfaces.
The H$_2$ diffusion barriers in the shallow, middle and deep silicate potential sites are 203 K, 308 K 
and 435 K respectively in model M07.    
Other than the H$_2$ diffusion barriers in silicate sites, parameters for model M07 are the same as these for model M02. Fig.~\ref{fig10} shows the 
recombination efficiency as a function of T predicted by the test model M07 and the reference model M01. 
We can see that H$_2$ can be efficiently synthesized ($\eta \geq 0.5$) 
when the grain temperatures are between 10 K and 14 K in model M07, 
which agrees well with model M01 predictions.
So, lower H$_2$ diffusion barriers can also enhance recombination efficiency in diffuse clouds. 
However, to the best our knowledge, the H$_2$ diffusion barriers on silicate surfaces are poorly known.
More work should be done to gauge this energy value in order to better understand H$_2$ formation in diffuse clouds.

Since the classical work by \citet{Hollenbach1971b}, models that adopted 
grain surfaces whose binding sites have various desorption energies and diffusion barriers 
have been successfully used to enhance the H atom recombination efficiency at higher temperatures.
The key to the success at higher temperatures is the existence of stronger H atom binding sites in these models.
Ironically, these stronger sites may become much weaker H atom binding sites after H$_2$ molecules occupy them.
Therefore, the success may not be as plausible as it seemed to be if we consider surface H$_2$ in models.
Our model results show that H$_2$ molecules must diffuse fast enough on grain surfaces 
so that they are not trapped on grain surfaces in order that
H$_2$ can be formed efficiently. 

We consider physisorption only in our models while 
H atoms can also be chemisorbed on silicate or carbon surfaces~\citep{Cazaux2004}. It is not clear whether models that involve 
chemisorption can explain H$_2$ formation in diffuse clouds or not if 
the existence of H$_2$ on grain surface is considered. 
More study should be done in order to unveil the mystery of H$_2$ formation in diffuse clouds. 

Hydrogen atoms can only be physisorbed on water ice surfaces~\citep{Wakelam2017a}.
Our translucent cloud model results further show that the conversion from physisorbed H atoms to H$_2$ molecules 
can be very slow on grain surfaces if the diffusion barrier of H$_2$ on water ice surface is large enough. 
Therefore, this diffusion barrier could be important for the molecular evolution of translucent clouds.
However, because surface chemical reactions that involve H$_2$ typically have a large reaction barrier, 
this diffusion barrier is not well studied so far. More rigorous research
should be done to gauge this energy value.  

\section*{Acknowledgments}
This work was funded by the National Natural Science Foundation of China (grant Nos. 11973099, 11973075 and 12173023).
We thank our referee for constructive comments to improve the quality
of the manuscript. We would like to thank Professor Laura Rowe for her help with proofreading and scientific English
writing. We thank Di Li for helpful discussions.

\section*{Data availability}
No new data were generated or analysed in support of this research.


\bsp

\label{lastpage}

\end{document}